\documentclass[twocolumn,showpacs,preprintnumbers,amsmath,amssymb]{revtex4}

\usepackage{graphicx}
\usepackage{dcolumn}
\usepackage{bm}

\newcommand{\bra}[1]{\langle#1\rvert}
\newcommand{\ket}[1]{\lvert#1\rangle}

\renewcommand{\vec}[1]{\bm{\mathrm{#1}}}

\expandafter\ifx\csname
 natexlab\endcsname\relax\def\natexlab#1{#1}\fi
 \expandafter\ifx\csname bibnamefont\endcsname\relax
   \def\bibnamefont#1{#1}\fi
 \expandafter\ifx\csname bibfnamefont\endcsname\relax
   \def\bibfnamefont#1{#1}\fi
 \expandafter\ifx\csname citenamefont\endcsname\relax
   \def\citenamefont#1{#1}\fi
 \expandafter\ifx\csname url\endcsname\relax
   \def\url#1{\texttt{#1}}\fi
 \expandafter\ifx\csname
 urlprefix\endcsname\relax\def\urlprefix{URL }\fi
 \providecommand{\bibinfo}[2]{#2}
 \providecommand{\eprint}[2][]{\url{#2}}

\begin{document}

\title{Quantum storage on subradiant states in an extended atomic ensemble}

\author{Alexey Kalachev}
\email{kalachev@kfti.knc.ru} \affiliation{Zavoisky
Physical-Technical Institute of the Russian Academy of Sciences,
Sibirsky Trakt 10/7, Kazan, 420029, Russia
}%

\date{\today}

\begin{abstract}
A scheme for coherent manipulation of collective atomic states is
developed such that total subradiant states, in which spontaneous
emission is suppressed into all directions due to destructive
interference between neighbor atoms, can be created in an extended
atomic ensemble. The optimal conditions for creation of such
states and suitability of them for quantum storage are discussed.
It is shown that in order to achieve the maximum signal-to-noise
ratio the shape of a light pulse to be stored and reconstructed
using a homogeneously broadened absorbtion line of an atomic
system should be a time-reversed regular part of the response
function of the system. In the limit of high optical density, such
pulses allow one to prepare collective subradiant atomic states
with near flat spatial distribution of the atomic excitation in
the medium.
\end{abstract}


\maketitle

\section{\label{sec:level1}Introduction}
The use of photons as quantum information carriers involves the
elaboration of effective quantum memory devices which are able to
write, store and reconstruct single photon quantum states of the
electromagnetic field. A promising approach to optical quantum
state storage uses the interaction of single photons with
optically dense media. The current activities focus on electromagnetically induced transparency (EIT)
\cite{FYL_2000,FL_2002,ZMKRWS_2002,CMJLKK_2005,DAMFZL_2005},
stimulated Raman absorption \cite{KMP_2000} and photon echo
\cite{KM_1993,MK_2001,MTH_2003,KTG_2005,NK_2005,MSG_2006,SSAG_2006}
phenomena. Besides, there is a successful experimental
demonstration of quantum memory for multi-photon quantum states
using off-resonant interaction of light with spin polarized atomic
ensembles \cite{JSCFP_2004}. In \cite{KS_2005,KK_2006} a scheme of
coherent manipulation of collective atomic states was developed
such that superradiant states of the atomic system can be
converted into subradiant ones and vice versa and possible
applications of such a scheme for optical quantum-state storage
were discussed. The advantage of the scheme is that storage and
retrieval of a single-photon state may in principle be implemented
by means of phase modulators only \cite{KK_2006}, i.e. without any
additional control fields or pulses acting on the atomic system.
On the other hand, the rate of collective spontaneous emission in
this case can be suppressed only for a few collective modes, which
means that storage time is limited by incoherent spontaneous
emission into other modes. In this paper the scheme is developed
such that total subradiant states, in which spontaneous emission
into all directions is suppressed, can be created in an extended
atomic system. Such an approach involves using homogeneously
broadened absorption lines for storage and retrieval of
information in an optically dense medium. In \cite{KK_2006} it was
pointed out that for high efficiency of such a quantum memory the
time shape of a single photon wave packet to be stored should be
equal to the time-reversed response function of the optically
dense medium. Then the shape of the emitted photon proves to be a
time-reversed replica of the initial one. In the present paper
this statement is discussed quantitatively and it is shown that
such pulses are optimal in the context of signal-to-noise ratio.
It should be noted here that the connection between optimal photon
storage and time reversal has been made in
\cite{MK_2001,KTG_2005} in considering storage on inhomogeneously
broadened transitions. A comprehensive analysis of the question
was presented recently in \cite{GAFSL_2006,GALS_2006}, where
optimal pulse shapes were derived providing the maximum efficiency
for different approaches to pulse storage. The results obtained
here do not contradict those presented in
\cite{GAFSL_2006,GALS_2006}, which will be discussed below in
detail, but involve another point of view, namely the
maximization of signal-to-noise ratio. Besides, instead of a
general iteration procedure which was used by authors of these
papers we use a more direct approach to the problem, where the
main features of the proposed scheme are explicitly taken into
account: the absence of a control field, homogeneous broadening of
an absorption line and forward retrieval. Our approach is based on
the theory of matched filters \cite{G_1953}, which allows us to
write down the explicit expressions for the optimal pulse shape
and total efficiency of quantum memory in a straightforward
manner.

The paper is organized as follows. In Sec.~II, we present a scheme
for coherent manipulation of collective atomic states that enables
the creation of total subradiant states in an extended atomic
system. In Sec.~III, the basic equations describing propagation of
single-photon wave-packets in an optically dense atomic medium are
introduced and the pulse shape which maximizes signal-to-noise
ratio upon read-out is determined.

\section{Quantum storage on subradiant states}
\begin{figure}
\includegraphics[width=8.6cm]{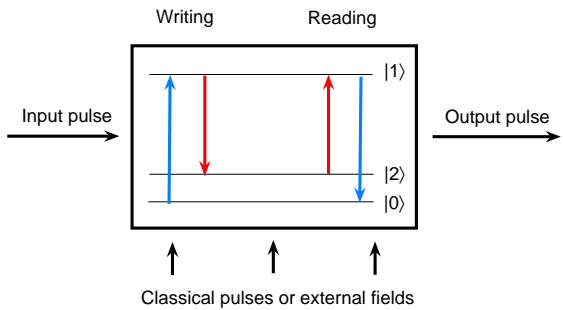}
\caption{\label{fig:scheme} (Color online) General scheme of
quantum memory device based on a three-level extended atomic
ensemble, collective states of which are controlled by an external
electric field or classical pulses.}
\end{figure}

In addition to the implementation schemes proposed in
\cite{KK_2006} we consider here another simple procedure for
writing and reconstructing single-photon states of light using
subradiant states. Consider an extended system of identical
three-level atoms forming an optically dense resonant medium (see
Fig.~\ref{fig:scheme}). We assume that the atoms are not moving
as, for example, impurities embedded in a solid state material. In
this case we may consider them to be distributed regularly in
space with an interatomic distance $a$ along some axis, say $x$.
Moreover, since the parity of energy states in such a system is
usually indefinite (due to the low symmetry of impurity sites), we
suppose that all transitions are dipole allowed.
Assume that a single-photon wave packet which is resonant to the
transition $\ket{0}\to\ket{1}$ propagates through the medium,
which has a phase relaxation time much longer than the duration of
the photon. At some moment of time the probability of finding the
medium in the excited state and the field in the vacuum state is
maximum and at this moment of time the atomic system can be
subjected to a short coherent $\pi$-pulse at the frequency of the
transition $\ket{1}\to\ket{2}$, which corresponds to the writing
of information. This step is typical of different quantum storage
techniques involving a $\Lambda$-type medium with inhomogeneously
broadened transitions \cite{MK_2001} as well as homogeneously
broadened ones \cite{GAFSL_2006}. Now, the spatial distribution of
phase of the atomic state $\ket{2}$ in the medium that results
from the excitation is described by a wave vector
$\vec{k}-\vec{k}_w$, where $\vec{k}$ and $\vec{k}_w$ are wave
vectors of the incoming photon and $\pi$-pulse (writing pulse),
respectively. If we put $|\vec{k}-\vec{k}_w|=\pi/a$, then the
excited states $\ket{2}$ of adjacent atoms along the axis $x$ will
be opposite in phase, provided that the vector $\vec{k}-\vec{k}_w$
is directed along $x$. Consequently, if the wavelength of the
transition $\ket{2}\to\ket{0}$, $\lambda_{20}$, satisfies the
condition $2\pi/\lambda_{20}<\frac{1}{2}|\vec{k}-\vec{k}_w|$ ---
i.e., $\lambda_{20}>4a$, then at least $2^3=8$ atoms prove to be
located in the volume $({\lambda_{20}}/{2})^3$, forming two equal
groups with opposite phases of the state $\ket{2}$. As a result, a
subradiant state is created, the rate of spontaneous emission of
photons from which is suppressed for all directions. In this state
the quantum storage is possible during times which may be much
longer than the state-$\ket{2}$ radiative lifetime of a single
atom. In order to read out the information it is necessary to
apply a short $\pi$-pulse (reading pulse) with the wave vector
$\vec{k}_r=\pm\vec{k}_w$ to the transition $\ket{2}\to\ket{1}$.
The signs $+$ and $-$ correspond to forward and backward retrieval.

Clearly the most promising materials for the creation of the
subradiant states are those in which homogeneous linewidth,
$\Gamma_h$, of the transition $\ket{2}\to\ket{0}$ is mainly
determined by the spontaneous relaxation of the excited state.
Linewidths approaching the limit $(\pi\Gamma_h)^{-1}\equiv
T_2=2T_1$, where $T_2$ is the phase memory time and $T_1$ is the
population life-time, can be seen in materials where all other
dephasing processes have been minimized, such as in
rare-earth-doped $\text{Y}_2\text{SiO}_5$ \cite{M_2002}, where a
$T_2$ of several ms has been observed. The difference
$1/T'_2=1/T_2-1/2T_1$, corresponding to such dephasing processes,
determines the life-time of the subradiant state, which may be an
order larger than $T_1$ at cryogenic temperatures. For example,
for the ${}^1D_2(1)-{}^3H_4(1)$ transition in
$\text{Pr}^{3+}:\text{Y}_2\text{SiO}_5$ we have $T_1=0.222$~ms and
$T_2=0.377$~ms \cite{ECM_1995}, therefore $T'_2=2.5$~ms. It should
be noted that in such materials at low temperatures the phase
memory of the hyperfine transitions, which are usually assumed to
be used for storage in a $\Lambda$-type medium, is limited by the
same processes that cause decoherence on the optical transitions,
most notably the magnetic interaction with spins in the host
material. Therefore, one can expect that the subradiant state
life-time is of the order of the hyperfine coherence time.
Increasing the latter by applying, for example, a static magnetic
field to the crystal implies increasing the former to the same
extent. On the other hand, the frequency of the transition
$\ket{2}\to\ket{0}$ may be much larger than $10^1-10^2$~MHz
typical of hyperfine transitions. Taking the interatomic distance,
$a=1.5\lambda_{10}$, we obtain $\lambda_{20}>6\lambda_{10}$, which
corresponds to the frequency $\omega_{20}<\omega_{10}/6$.

As for homogeneous absorption lines, the technique of preparing of
narrow absorbing peaks on a non-absorbing background, i.e.
isolated spectral features corresponding to a group of ions
absorbing at a specific frequency, in rare-earth-metal-ion-doped
crystals
\cite{PSM_2000,SPMK_2000,NROCK_2002,SLLG_2003,NRKKS_2004,RNKKS_2005}
can be very useful. Such specific structures can be created as
follows. First, spectral pits, i.e. wide frequency intervals
within the inhomogeneous absorption profile that are completely
empty of all absorption, are created using hole-burning
techniques. Then narrow peaks of absorption are created by pumping
ions absorbing within a narrow spectral interval back into the
emptied region. The peaks can have a width of the order of the
homogeneous linewidth, if a laser with a sufficiently narrow
linewidth is used for the preparation. Moreover, using two
non-collinear laser pulses instead of a single one it is possible
to prepare a periodic structure with a necessary spatial period
$a$, since the atoms will be pumped mainly within antinodes of the
laser field.

Finally, it should be noted that the total subradiant states in a
macroscopic atomic ensemble can be created in principle by
significantly changing the refractive index of the medium. There are
many proposals aimed at the enhancement of refractive index with
vanishing absorption based on quantum interference effects which
trace back to the works by Scully and colleagues (see
\cite{SZ_1997} and references therein). Bearing them in mind we
can consider the following procedure. Let the initial value of the
refractive index of the host material be enhanced, so that the
wavelength $\lambda_{10}$ satisfies the condition
$a=3\lambda_{10}/2$. Then rather than apply a short coherent
$\pi$-pulse, we can reduce the refractive index at least by a
factor of 6, so that $\lambda_{10}$ becomes larger than $4a$ as
in the previous case. As a result, a total subradiant state is
created. By combining both techniques (applying the $\pi$-pulse
and changing the refractive index) it is possible to lift the
restriction $\omega_{20}<\omega_{10}/6$.

\section{Optimization of signal-to-noise ratio}
The efficiency of the scheme considered above as well as of those
considered in \cite{KK_2006} depends strongly on the possibility
of full (at some moment of time) photon absorption in the medium
with a homogeneously broadened resonant transition and subsequent
emission of the photon in the same direction. This is possible
only for a specific (optimal) time shape of the pulses to be
stored. The optimal pulse shape which leads to the maximum
efficiency of quantum storage may be found numerically using an
iteration optimization procedure \cite{GAFSL_2006,GALS_2006}. Here
we consider the pulse shape which leads to the maximum peak value
of the retrieved signal, given the energy of the pulse. Such pulses
may be referred to as optimal in respect to the signal-to-noise ratio,
the criterion generally employed in communications. It will be
shown below that such pulses create almost uniform distribution
of atomic excitation in the medium at some moment of time,
corresponding to the end of the incoming pulse. This moment is
optimal for application of the short writing $\pi$-pulse creating
a subradiant state. On the other hand, upon read out the emitted
field proves to be the time-reversed replica of the initial field,
which is the characteristic feature of the optimization
\cite{GAFSL_2006,GALS_2006}. Such a regime may be useful, for
example, for a long-distance quantum communication using quantum
repeaters \cite{BDCZ_1998}, when the qubits are only stored and
recalled once before being measured. Assuming, for example, that
time-bin qubits are used for carrying the information, each of
them should be a superposition of well separated wave packets of
the optimal shape, which provides high efficiency and fidelity of
quantum memory devices. The latter characterizes reproducibility
of the relative phase and amplitude ratio of time-bin
single-photon pulses during storage and retrieval.

In the simplest case, of an additive white noise, the maximization
of the signal-to-noise ratio reduces to that of the amplitude of the
output signal at some moment of time, given the energy of the
input signal and transfer function of the medium. Although such a
procedure is performed usually in the context of classical signals
\cite{G_1953,CB_1967,P_1995}, it works exactly the same when a
single photon should be detected at the output of a memory device.
The only difference is that the amplitude and intensity of
classical light are replaced by the photon probability amplitude
density and photon probability density, respectively. A relevant
situation may be the detection of single photons amid broadband
background light in free-space communication setups
\cite{GRTZ_2002}. The maximization of single-photon probability
density at some moment of time means shortening of the
single-photon wave packet to be stored and recalled using a given
absorbtion line. This allows one to minimize the time windows
which are necessary for the writing and read out of information
and consequently to minimize the probability of detection of stray
photons instead of information carriers. The same argument is true in
the context of noise due to the dark counts of single-photon
detectors.

\subsection{Basic equations}
Consider a system of $N\gg 1$
identical two-level atoms, with positions $\vec{r}_j$
($j=1,\ldots,N$) and resonance frequency $\omega_0$, interacting
among themselves and with the external world only through the
electromagnetic field. We are interested in the interaction of the
atomic system with a single-photon wave packet. In a
one-dimensional light propagation model it is usually assumed that
the excitation volume may be approximated by a cylinder with the
cross section $S$ and the length $L$, the Fresnel number of the
excitation volume $F=S(L\lambda)^{-1}\geq 1$, a single-photon wave
packet propagates in the $z$-direction, and the wave front of the
packet is planar inside the excitation volume. Let us divide the
medium into $n$ identical slices of mean position $z_p=pL/n$
($p=1,2,\ldots,n$). The length of each slice $\Delta z$ is large
compared to the wavelength $\lambda=2\pi c\omega_0^{-1}$, but
small compared to $L$. We assume that each slice contains a large
number $N_p=N/n$ of atoms, but has a small optical density
$\alpha\Delta z\ll 1$, where $\alpha$ is a resonant absorption
coefficient. Therefore, hereafter we assume that slowly
time-varying envelopes of the field and atomic probability
amplitudes are constant in each slice and consider
"coarse-grained" functions on coordinate $z$. Besides, we assume
that the time of propagation of photon through the system $L/c$ is
negligibly short compared to the evolution time of the slowly
time-varying envelopes.

Let us denote the ground and excited states of $j$th atom by
$\ket{0_j}$ and $\ket{1_j}$. The Hamiltonian of the system, in the
interaction picture and rotating-wave approximation, reads
\begin{equation}\label{Ham}
H=\sum_{j,\vec{k},s}\hbar g_{\vec{k},s}^\ast b_j^\dag
a_{\vec{k},s}{\,e}^{i\vec{k}\cdot\vec{r}_j}
{\,e}^{i(\omega_0-\omega)t}+\text{H.c.}
\end{equation}
Here $
g_{\vec{k},s}=\frac{i}{\hbar}\left(\frac{\hbar\omega}{2\varepsilon_0
V}\right)^{1/2}(\vec{d}\cdot\vec{\varepsilon}_{\vec{k},s})$  is
the atom-field coupling constant, $b_j=\ket{0_j}\bra{1_j}$ is the
atomic transition operator, $a_{\vec{k},s}$ is the photon
annihilation operator in the radiation field mode with the
frequency $\omega=kc$ and polarization unit vector
$\vec{\varepsilon}_{\vec{k},s}$ ($s=1,2$), $V$ is the quantization
volume of the radiation field (we take $V$ much larger than the
volume of the atomic system), $\vec{d}$ is the dipole moment of
the atomic transition. For the sake of simplicity we assume that
the vectors $\vec{\varepsilon}_{\vec{k},s}$ and $\vec{d}$ are
real.

First, consider the system of a slice of atoms and electromagnetic
field. For each slice with coordinate $z_p$ we can define the
following collective atomic operators:
\begin{equation}
R_p=\sum_{j=1}^{N_p} b_j{\,e}^{- i\vec{k}_0\cdot\vec{r}_j},
\end{equation}
where $\vec{k}_0$ is directed along $z$-axis and
$|\vec{k}_0|=\omega_0/c$, and the general form of the state of the
system can be written as
\begin{equation}\label{psi_0}
\ket{\psi(t,z_p)}=\sum_{\vec{k},s}f_{\vec{k},s}(t,z_p)\ket{0}\ket{1_{\vec{k},s}}+
c(t,z_p)\ket{1}\ket{\text{vac}}
\end{equation}
with normalization condition
$\sum_{\vec{k},s}|f_{\vec{k},s}(t,z_p)|^2+|c(t,z_p)|^2=1$,  where
$\ket{0}=\ket{0_1,0_2,\ldots,0_{N_p}}$ is the ground state of the
slice's atomic system, $\ket{\text{vac}}$ is the vacuum state of
the radiation field,
$\ket{1_{\vec{k},s}}=a^\dag_{\vec{k},s}\ket{\text{vac}}$ and
$\ket{1}=N_p^{-1/2}R_p^{\dag}\ket{0}$. It should be noted that the
normalization condition right after Eq.~(\ref{psi_0}) is written
for the system consisting of only one slice and the emf. This
normalization condition will not used when considering the whole
atomic system.

Substituting Eqs.~(\ref{Ham}) and (\ref{psi_0}) in the
Schr\"{o}dinger equation we obtain
\begin{align}
\frac{\partial f_{\vec{k},s}(t,z_p)}{\partial
t}=&-ig_{\vec{k},s}\sqrt{N_p}\phi(\vec{k}_0-\vec{k})
c(t,z_p){\,e}^{-i(\omega_0-\omega)t},\label{Eq1}\\
\frac{\partial c(t,z_p)}{\partial t}=&-i\sqrt{N_p}\nonumber\\
&\times\sum_{\vec{k},s}g^\ast_{\vec{k},s}\phi^\ast(\vec{k}_0-\vec{k})
f_{\vec{k},s}(t,z_p){\,e}^{i(\omega_0-\omega)t},\label{Eq2}
\end{align}
where $\phi(\vec{x})=N_p^{-1}\sum_{j}{\exp}(i\vec{x}\cdot\vec{r}_j)$
is the diffraction function.

The photon density for the incoming wave packet at the slice reads
\begin{equation}
F_\text{in}(t,z_p)=\frac{1}{L^{3/2}}\sum_{\vec{k},s}
f_{\vec{k},s}(-\infty,z_p){\,e}^{i(\omega_0-\omega)t},
\end{equation}
and for the emitted radiation we have the analogous equation with
$F_\text{in}(t,z_p)$ and $f_{\vec{k},s}(-\infty,z_p)$ replaced by
$F(t,z_p)$ and $f_{\vec{k},s}(t,z_p)$, respectively. Then the
solution of Eqs.~(\ref{Eq1}) and (\ref{Eq2}) may be written as
\begin{align}
c(t,z_p)={}&c(-\infty,z_p){\,e}^{-(N_p\mu +1) t/2T_1}\nonumber\\
&-\sqrt{\frac{N_p\mu}{T_1}}\int_0^\infty d\tau
F_\text{in}(t-\tau,z_p){\,e}^{-(N_p\mu
+1)\tau/2T_1},\label{Solution_0a}
\end{align}
\begin{equation} F(t,z_p)=F_\text{in}(t,z_p)+\sqrt{\frac{N_p\mu}{T_1}}\,c(t,z_p),\label{Solution_0b}
\end{equation}
Here $\mu=3\lambda^2(8\pi S)^{-1}$ is a geometrical factor
\cite{RE_1971}, which describes the result of the integration
\begin{equation}
\int
d\Omega_{\vec{k}}\sum_s(\vec{d}\cdot\vec{\varepsilon}_{\vec{k},s})^2
\phi^2(\vec{k}_0-\vec{k})=\frac{8\pi}{3}\left(\mu+\frac{1}{N_p}\right)
d^2
\end{equation}
for identical dipole moments oriented perpendicular to the $z$-axis,
and
\begin{equation}\label{tau_R}
\frac{1}{T_1}=\frac{1}{4\pi\varepsilon_0}\frac{4d^2\omega_0^3}{3\hbar
c^3}.
\end{equation}

If we consider the case when $c(-\infty,z_p)=0$ and substitute
Eq.~(\ref{Solution_0a}) into (\ref{Solution_0b}), we obtain a
solution for superradiant resonant forward scattering of photons by
an optically thin atomic medium \cite{BC_1969,C_1970}:
\begin{align}\label{Fscat}
F(t,z_p)={}&F_\text{in}(t,z_p)\nonumber\\&-b(\Delta
z)\int_0^\infty d\tau F_\text{in}(t-\tau,z_p){\,e}^{-\tau/T_2}.
\end{align}
where $b(x)= \alpha x/2T_2$, $T_2=2T_1$, $\alpha=4\mu N_p/\Delta z$ is the
resonant absorption coefficient and we have omitted $N_p\mu \ll 1$
from the exponential.

The solution (\ref{Fscat}) can be written in terms of the
impulse-response function or transfer function of the slice. If we
define
\begin{equation}
F_\text{in}(\omega)=\frac{1}{\sqrt{2\pi}}\int_{-\infty}^{\infty} dt
F_\text{in}(t)\,e^{i\omega t},
\end{equation}
then
\begin{align}\label{LinearSystem}
F(t,z_p)&=\int_{-\infty}^{\infty}d\tau
F_\text{in}(\tau,z_p)H(t-\tau)\nonumber\\
&=\frac{1}{\sqrt{2\pi}}\int_{-\infty}^{\infty}d\omega
F_\text{in}(\omega,z_p)H(\omega)\,e^{-i\omega t},
\end{align}
where
\begin{align}
H(t)&=\delta(t)-b(\Delta
z)\,\theta(t)\,e^{-\frac{t}{T_2}},\label{Ht}\\
H(\omega)&=1-b(\Delta z)\frac{i}{\omega+{i}/{T_2}}.\label{Homega}
\end{align}
Here $\theta(t)$ is equal to 0 for $t<0$, 1 for $t>0$ and $1/2$
for $t=0$.

Now we return to the case of an optically dense medium considered
as a sequence of optically thin slices, each of them
characterized by the impulse-response (\ref{Ht}) or transfer
(\ref{Homega}) function. In this case we can consider the quantity
$c'(t,z)=\lim_{\Delta z\to 0}c(t,z_p)/\sqrt{\Delta z}$ as a
probability amplitude density and assume that $z\in[0,L]$. For an
optically thick medium the transfer function becomes
\begin{align}\label{HomegaThick}
H(\omega,L)&=
\lim_{n\to\infty}\left(1-b\left(\frac{L}{n}\right)\frac{i}{\omega+{i}/{T_2}}\right)^n\nonumber\\
&=\exp\left(-b(L)\frac{i}{\omega+{i}/{T_2}}\right).
\end{align}
By expanding the exponential in Eq.~(\ref{HomegaThick}) in power
series and performing the Fourier transformation we obtain the
following impulse-response function of a resonant medium with
arbitrary optical density \cite{C_1970}
\begin{equation}\label{HtThick}
\begin{aligned}
H(t,L)&=\delta(t)-\Phi(t),\\
\Phi(t)&=b(L)\frac{J_1(2\sqrt{b(L)t})}{\sqrt{b(L)t}}\,\theta(t)\,e^{-\frac{t}{T_2}}.
\end{aligned}
\end{equation}
Here $J_1(x)$ is the Bessel function of the first kind. Taking
into account the coordinate dependence $F(t,z)$ for the optically
thick sample, Eq.~(\ref{LinearSystem}) should be written as
\begin{align}\label{LinearSystem2}
F(t,L)&=\int_{-\infty}^{\infty}d\tau
F_\text{in}(\tau,0)H(t-\tau,L)\nonumber\\&=\frac{1}{\sqrt{2\pi}}\int_{-\infty}^{\infty}d\omega
F_\text{in}(\omega,0)H(\omega,L)\,e^{-i\omega t}.
\end{align}

\subsection{Optimal pulse shape}
\begin{figure}
\includegraphics[width=8.6cm]{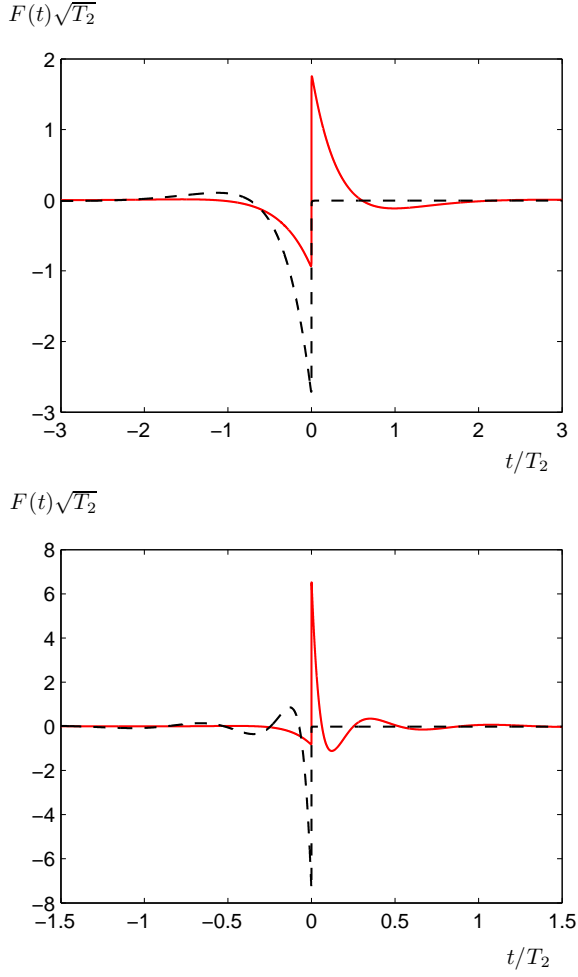}
\caption{\label{fig:response} (Color online) The amplitudes
$F_\text{in}(t,0)$ (dashed line) and $F(t,L)$ (solid line) as
functions of time $t$ for $\alpha L=10$ (above) and $\alpha L=100$
(below).}
\end{figure}

First, find a
shape of single photon wave packets to be stored, which maximizes
the signal-to-noise ratio (SNR) at the retrieval. In the theory of
linear filters \cite{G_1953} it is well known that the maximum SNR
is achieved for so called matched filters, the impulse-response
function of which is a time-reversed replica of an input
(detected) signal, provided that the signal is read through a
white noise. In that case the maximum peak value of the output
signal, given energy of the input signal, is achieved at some
moment of time. In the present case of a homogeneously broadened
absorption line, it is necessary to obtain the highest possible
peak value of the photon density
\begin{multline}\label{F2}
|F(t,L)|^2=\left[\int_{-\infty}^{\infty}d\tau
F_\text{in}(\tau,0)H(t-\tau,L)\right]^2\\
=F_\text{in}^2(t,0)-2F_\text{in}(t,0)\int_{-\infty}^{\infty}d\tau
F_\text{in}(\tau,0)\Phi(t-\tau,L)\\
+\left[\int_{-\infty}^{\infty}d\tau
F_\text{in}(\tau,0)\Phi(t-\tau,L)\right]^2
\end{multline}
at some moment of time $t\geq 0$, assuming that the incoming pulse
terminates at $t=0$ and $F_\text{in}(t,0)$ is a real function. The
field generated after the moment $t=0$ is determined only by the
last term in Eq. (\ref{F2}), the maximum peak value of which is
achieved at the moment $t=0$ provided that
$F_\text{in}(\tau,0)\propto \Phi(-\tau,L)$, which follows from
the Cauchy---Bunyakowsky---Schwartz inequality. So, taking into
account Eq.~(\ref{HtThick}) we suggest the following single photon
wave packet
\begin{align}\label{Fin}
F_\text{in}^{L}(t,0)&=-A(L)\Phi(-t)\nonumber\\
&=\frac{1}{\sqrt{2\pi}}\int_{-\infty}^{\infty} d\omega
\frac{A(L)}{\sqrt{2\pi}}\left[H(-\omega,L)-1\right]\,e^{-i\omega t}
\end{align}
as an optimal one from the standpoint of the SNR criterion. Here
$A(L)=[b(L)(1-g(L))]^{-1/2}$ is the normalization constant,
$g(x)=e^{-\alpha x/2}(I_0(\alpha x/2)+I_1(\alpha x/2))$, $I_n(x)$
is the modified Bessel function of the first kind. The pulse
begins at $t=-\infty$ and terminates at $t=0$, but in fact almost
all energy is concentrated in several last oscillations of its
amplitude. Substituting Eq.~(\ref{Fin}) in
Eq.~(\ref{LinearSystem2}) and using Eq.~(\ref{HomegaThick}) or
Eq.~(\ref{HtThick}) we obtain
\begin{equation}\label{Solution_main}
F(t,L)=\gamma(t,L)-F_\text{in}^{L}(-t,0),
\end{equation}
where
\begin{align}
\gamma(t,L)=&A(L)\sqrt{\frac{\pi|t|}{2T_2}}\sum_{m=1}^\infty
\frac{(\alpha L/2T_2)^m}{m!(m-1)!}|t|^{m-1}\nonumber\\
&\times[I_{m-1/2}(|t|/T_2)-I_{-m+1/2}(|t|/T_2)]
\end{align}
is a near bell-shaped function, which in the case $\alpha L\gg 1$ is
approximated by
\begin{equation}\label{GammaThick}
\gamma(t,x)=-\sqrt{b(x)}\frac{g(x)}{\sqrt{1-g(x)}}\exp(-|t|\sqrt{\alpha
x}/T_2),
\end{equation}
and in the limit $\alpha L\to 0$ takes the form
$\gamma(t)=-\sqrt{2/T_2}\exp(-|t|/T_2)$. In the last case the
optimal shape of the incoming pulse becomes
$F_\text{in}(t)=-\sqrt{2/T_2}\,\theta(-t)\exp(t/T_2)$, therefore
from Eq.~(\ref{Solution_main}) it follows that the superradiant
forward scattering is negligible as expected. In the opposite case
$\alpha L\to \infty$ the role of the function $\gamma(t,L)$ on the
right hand side of Eq.~(\ref{Solution_main}) tends to zero since
$\int_{-\infty}^{\infty}|\gamma(t,L)|^2\,dt=2(\pi\sqrt{\alpha
L})^{-1}$, so that no field goes through the medium until $t\geq
0$ (see Fig.~\ref{fig:response}).
At the moment $t=0$ corresponding to the end of the incoming pulse the
atomic system starts to emit the outcoming pulse
$F(t,L)=-F_\text{in}^{L}(-t,0)$ which is the time-reversed (and
opposite in phase) replica of the input pulse. Thus the moment $t=0$
is optimal for instantaneous creation of a subradiant state and
mapping a single-photon state onto the collective atomic one.

Now, consider the efficiency of the proposed quantum memory
scheme, defined by
\begin{equation}\label{Eff}
\mathcal{E}\equiv \frac{\int_0^\infty
|F_\text{out}(t)|^2dt-\int_{-\infty}^0
|F_\text{out}(t)|^2dt}{\int_{-\infty}^0 |F_\text{in}(t)|^2dt}.
\end{equation}
The denominator is equal to unity by definition and the second
term in the numerator corresponds to the probability of the photon
loss due to its emission before the moment $t=0$, corresponding to
the end of the incoming pulse and creation of a subradiant state.
Substituting Eqs. (\ref{Fin}), (\ref{Solution_main}) and
(\ref{GammaThick}) we obtain
\begin{equation}\label{Eff2}
\mathcal{E}=1-\frac{4}{\sqrt{\pi\alpha L}}.
\end{equation}
As expected, the efficiency tends to unity in the limit $\alpha
L\to\infty$ according to a square-root law, which is typical of
propagation effects in homogeneously broadened resonant systems.
The main source of the photon loss is its leakage through the
sample and incoherent emission in transverse modes during the
whole process of writing and read-out. It should be noted also
that with increasing of $\alpha L$ the duration of the optimal
single-photon wave-packet decreases.

Finally, we consider the spatial distribution of probability
amplitude density $c'(t,z)$ at $t=0$. From Eq.
(\ref{LinearSystem2}) it follows that the incoming pulse
(\ref{Fin}) after propagation of distance $z\leq L$ takes the
form:
\begin{align}\label{Field_inside}
F(t,z)=&\frac{A(L)}{A(z)A(L-z)}\int_{-\infty}^{\infty}d\tau\,\gamma(t-\tau,z)F_\text{in}^{L-z}(\tau,0)\nonumber\\
&+\frac{A(L)}{A(z)}\gamma(t,z)+
\frac{A(L)}{A(L-z)}F_\text{in}^{L-z}(t,0)\nonumber\\
&-\frac{A(L)}{A(z)}F_\text{in}^{z}(-t,0).
\end{align}
On the other hand from Eqs.~(\ref{Solution_0a}) and
(\ref{LinearSystem2}) we can write
\begin{align}\label{c_z}
c'(0,z)=&\sqrt{\frac{\alpha}{2T_2}}\int_{-\infty}^\infty d\omega\,
F_\text{in}^L(\omega,0)H(\omega,z)\nonumber\\&\times\frac{i}{\sqrt{2\pi}(\omega+i/T_2)}=-\sqrt{\frac{2T_2}{\alpha}}\frac{\partial
F(0,z)}{\partial z}.
\end{align}
Using the approximation (\ref{GammaThick}) we obtain from
Eq.~(\ref{Field_inside})
\begin{align}
F(0,z)=A(L)\Bigg[&-b(z)g(z)\exp\left(-\frac{b(L-z)T_2}{1+\sqrt{\alpha
z}}\right)\nonumber\\ &-\frac{b(L-z)}{2}+\frac{b(z)}{2}\Bigg],
\end{align}
therefore
\begin{equation}\label{c_z_2}
c'(0,z)=-\frac{A(L)\sqrt{b(L)}}{\sqrt{L}}\left[1-\frac{1}{\sqrt{\pi}}\exp\left(-\frac{\alpha(L-z)}{2\sqrt{\alpha
L}}\right)\right],
\end{equation}
provided that $\alpha L\gg 1$.

\begin{figure}
\includegraphics[width=8.6cm]{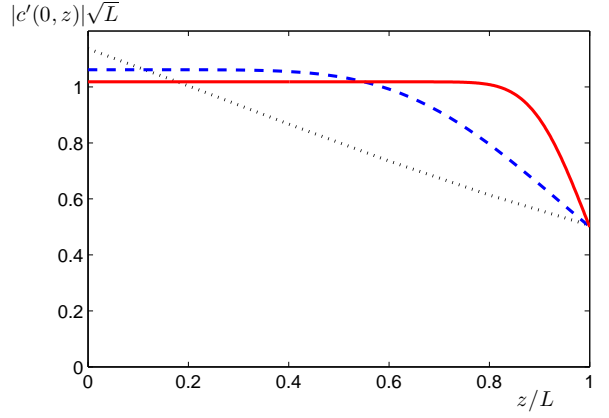}
\caption{\label{fig:amplitude} (Color online) The amplitude density
$|c'(0,z)|$ as a function of coordinate $z$ for $\alpha L=10$
(dotted line), $\alpha L=100$ (dashed line) and $\alpha L=1000$
(solid line). The results of numerical calculation using
Eq.~(\ref{c_z}).}
\end{figure}
We conclude that in the case of an optically thick sample the
energy of the incoming wave packet, having optimal pulse shape, is
distributed almost uniformly in the medium at the moment $t=0$
except for the far end of the sample ($z=L$) with the thickness
$\Delta z/L\approx {2\ln 2}/{\sqrt{\alpha L}}$. (see
Fig.~\ref{fig:amplitude}). The probability of photon absorption at
the moment $t=0$ is equal to
\begin{align}\label{p}
p_\text{abs}(t=0)&=\int_0^{L}|c'(0,z)|^2\,dz\nonumber\\&=
1-\frac{2}{\sqrt{\pi\alpha L}}+\frac{1}{\pi\sqrt{\alpha L}}.
\end{align}
The second term on the right hand side of Eq.~(\ref{p})
corresponds to the photon loss before the moment $t=0$, which is
half of the total photon loss probability in Eq.~(\ref{Eff2}).

When our results are compared with those of
\cite{GAFSL_2006,GALS_2006}, it is apparent that the pulse shape
(\ref{Fin}) does not provide the maximum of efficiency
(\ref{Eff}). The latter needs the maximization of the total
probability $\int_0^\infty dt\left[\int
d\tau\Phi(t-\tau,L)F_\text{in}(\tau,0)\right]^2$ instead of the
value $\left[\int d\tau\Phi(-\tau,L)F_\text{in}(\tau,0)\right]^2$.
As a result the discontinuities of the amplitude of the atomic
excitation at the edges of the sample arise and the error scales
as one over square root of optical depth (see Eq.~(\ref{Eff2})).
On the other hand, the pulse shape considered here gives the
maximum peak value of the probability density of the retrieved
single-photon wave packet. Numerics show that in the case $\alpha
L\gg 1$ the first burst of the output pulse contains about 90\% of
total photon probability, whereas that of the pulse optimized in
respect to the efficiency is only about 70\% having approximately
the same duration \cite{numerics}.

\section{Conclusion}
In summary, the quantum storage on a subradiant state in a
macroscopic atomic ensemble is analyzed, provided that on a
homogeneously broadened absorption line, no control field and
forward retrieval are used. In possession of the impulse-response
function of the sample it is possible to optimize the process of
photon storage and retrieval from the standpoint of
the signal-to-noise ratio and write down an explicit expression for
the optimal pulse shape, which may be useful in the context of its
experimental preparation. The light pulse to be stored should have
a shape which is a time-reversed replica of the impulse-response
function (to be more precise, of its regular part) of the atomic
system. At the moment of time corresponding to the end of the
absorbed pulse and the beginning of the emitted pulse the
probability of excitation in the medium is distributed almost
uniformly along the propagation direction. Therefore this is the
optimal moment of time for coherent manipulation of the collective
atomic state aimed at the capture of the photon \cite{KK_2006} or
preparation of the subradiant state in an extended atomic
ensemble. Obviously, the results obtained here can be easily
generalized for absorption systems with another impulse-response
function and for different kinds of noise, if the conditions for
the linear response theory are fulfilled.

\begin{acknowledgements}
The author would like to thank Stefan Kr\"{o}ll and Rustem
Shakhmuratov for useful comments and discussions. This work was
supported by the Program of the Presidium of RAS 'Quantum
macrophysics'.
\end{acknowledgements}


\end{document}